\documentclass[lettersize,journal]{IEEEtran}

\IEEEoverridecommandlockouts

\usepackage[cmex10]{amsmath}
\usepackage{amsthm}
\usepackage{amsmath}
\usepackage{algorithmic}
\usepackage{array}
\usepackage{bm,cite}
\usepackage{amssymb}
\usepackage[tight,footnotesize]{subfigure}
\usepackage{url,caption, color, enumerate, epsfig}

\begin{document}

\title{How to Attack and Defend NextG Radio Access Network Slicing with Reinforcement Learning}
\author{Yi Shi,~\IEEEmembership{Senior Member,~IEEE}, Yalin E. Sagduyu,~\IEEEmembership{Senior Member,~IEEE}, Tugba Erpek,~\IEEEmembership{Member,~IEEE}, \\
and M. Cenk Gursoy,~\IEEEmembership{Senior Member,~IEEE}
\thanks{This effort is supported by the U.S. Army Research Office under contract W911NF-17-C-0090. The content of the information does not necessarily reflect the position or the policy of the U.S. Government, and no official endorsement should be inferred.}
\thanks{Yi Shi is with Commonwealth Cyber Initiative, Virginia Tech, Arlington, VA 22203, USA. (e-mail: yshi@vt.edu)}
\thanks{Yalin E. Sagduyu and Tugba Erpek are with National Security Institute, Virginia Tech, Arlington, VA 22203, USA. (e-mail: ysagduyu@vt.edu; terpek@vt.edu)}
\thanks{M. Cenk Gursoy is with Syracuse University, Syracuse, NY 13244, USA. (e-mail: mcgursoy@syr.edu)}
}

\newcommand{\argmax}{\arg\!\max}

\maketitle

\begin{abstract}
In this paper, reinforcement learning (RL) for network slicing is considered in NextG radio access networks, where the base station (gNodeB) allocates resource blocks (RBs) to the requests of user equipments and aims to maximize the total reward of accepted requests over time. Based on adversarial machine learning, a novel over-the-air attack is introduced to manipulate the RL algorithm and disrupt NextG network slicing. The adversary observes the spectrum and builds its own RL based surrogate model that selects which RBs to jam subject to an energy budget with the objective of maximizing the number of failed 
requests due to jammed RBs. By jamming the RBs, the adversary reduces the RL algorithm's reward. As this reward is used as the input to update the RL algorithm, the performance does not recover even after the adversary stops jamming. This attack is evaluated in terms of both the recovery time and the (maximum and total) reward loss, and it is shown to be much more effective than benchmark (random and myopic) jamming attacks. Different reactive and proactive defense schemes (protecting the RL algorithm's updates or misleading the adversary's learning process) are introduced to show that it is viable to defend NextG network slicing against this attack.
\end{abstract}


\section{Introduction}

\subsection{Machine Learning for NextG Radio Access Network Slicing}

NextG offers major enhancements to the performance of cellular communications to meet the data rate demands of emerging applications such as virtual/augmented reality and Internet of Things. 
One key component of NextG communications is \emph{network slicing} in the \emph{radio access network} (RAN), which splits communication resources into virtual resource blocks (RBs). 
These RBs can be allocated dynamically to support different types of user applications.
These applications are categorized as enhanced Mobile Broadband (eMBB), massive machine-type communications (mMTC) and ultra-reliable low-latency communications (URLLC) based on throughput and latency requirements. Efficient and fast resource allocation by RAN slicing is critical  for near-real time RAN Intelligent Controller (Near-RT RIC). The details on resource allocation as part of RAN slicing are not defined yet in the 3GPP standards. To address this gap, research activities have focused on how the resources should be allocated as part of RAN slicing \cite{KaloxylosSurvey, Foukas, Ordonez,Rost, TommasoRAN}.

\emph{Machine learning} provides automated means to learn from data and optimize decision making for complex tasks. Supported by recent algorithmic and computational advances, \emph{deep learning} can operate on raw data without hand-crafted feature extraction and learn the underlying complex data representations. Therefore, deep learning has found rich applications in wireless communications such as waveform design, spectrum situational awareness, and wireless security \cite{WirelessDL}. Related to network slicing, deep learning was studied in \cite{NakaoML} for application and device specific identification and traffic classification problems, and  in \cite{ThantharateML} for management of network load efficiency and network availability. Instead of relying on the availability of training data, \emph{reinforcement learning} (RL) has emerged as a viable solution for NextG network slicing \cite{KooSlicingRL, LiSlicingRL, LiuSlicingRL, WangSlicingRL, Gursoy, CAMAD, CAMADJournal, Yasin, Suh22} such as learning from the NextG network performance and updating resource allocation decisions for network slicing.

In this paper, we consider a NextG base station, i.e., gNodeB, as the victim system that runs an RL algorithm (as an example, the \emph{Q-learning} algorithm) to dynamically allocate resources for NextG network slicing, where RBs are allocated to support downlink communications from the gNodeB to the user equipments (UEs). Each network slicing request from any UE is associated with user-centric priority (weight), throughput and latency (deadline) requirements (namely, the quality of experience (QoE)), and needs to be served for a specific duration.

\subsection{Adversarial Machine Learning based Attack on NextG Radio Access Network Slicing with Reinforcement Learning}

Due to the broadcast nature of wireless communications, an adversary can overhear and jam transmissions. As a consequence, the adversary can launch a \emph{jamming attack} on RBs. Separate from NextG network slicing, attacks on RL algorithms have been considered in \cite{Cenk1, Cenk2, Cenk3} for medium access with a jammer that can jam one channel over one time block only. In this paper, we consider allocation of potentially multiple channels to different users over a time horizon for the NextG network slicing problem.  If an RB is assigned to a network slicing request and is jammed by the adversary, this request cannot achieve the required QoE and is considered as a failure. The reward of this request becomes zero, i.e., the performance of the gNodeB is reduced under attack. Moreover, this reward is given as the input (along with the state) to the gNodeB's RL algorithm. 
Therefore, this algorithm is confused and will predict the existence of jamming attacks even if there is no attack. 
Thus, such a jamming attack not only affects the gNodeB's current performance but also affects its future performance even after the adversary stops jamming RBs. On the other hand, RL can recover from the attack over a period of time by collecting correct feedback once the attack stops and updating its algorithm. To measure the performance of this attack (in terms of its effect on NextG network slicing), we compute the \emph{recovery time}, which is the time period from when the jamming attack stops to when the gNodeB's performance is back to normal (i.e., to the level before the attack starts), as well as the maximum and total reduction in the RL algorithm's reward during the recovery time.

We impose the practical constraint that the adversary has limited transmit power and thus cannot jam all RBs due to its \emph{energy budget}. Then, the adversary needs to carefully select which RBs to jam with the objective of maximizing the impact of jamming on network slicing requests (namely, the number of failed network slicing requests). One potential attack strategy is \emph{myopic}, which aims to jam some RBs to maximize the instantaneous impact of the attack without consideration of future impact. This strategy 
cannot work well as an online algorithm in general.
Moreover, our results show that this rather simple strategy can be learned by the gNodeB's RL algorithm and thus its impact can be mitigated over time by the usual RL algorithm updates.

\begin{figure}
	\centering
	\includegraphics[width=0.6\columnwidth]{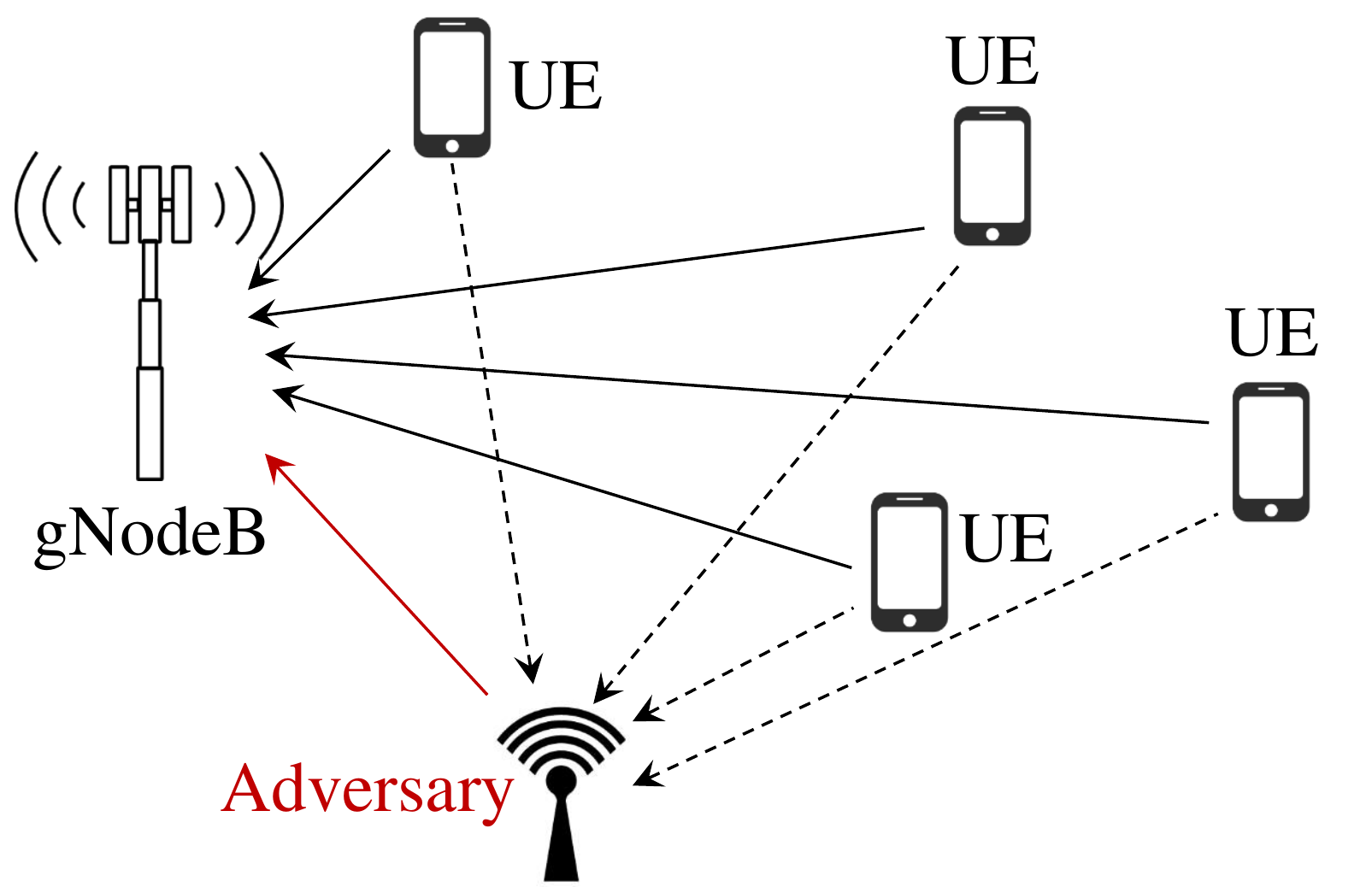}
	\caption{The interaction of the victim RL algorithm and the adversarial surrogate RL algorithm. The solid black lines from the UEs to the gNodeB represent the control messages sent for a failed network slicing request. The dashed lines from the UEs to the adversary represent the control messages for a failed network slicing request heard at the adversary. The adversary decides on its attack strategy based on its environment sensing results.}
	\label{fig:system}
\end{figure}

To maximize the impact of jamming the RBs, we pursue an \emph{adversarial machine learning} approach. 
Different types of attacks built upon adversarial machine learning have been studied in wireless communications \cite{Meets2021, Qian2021} such as exploratory (inference) attacks \cite{Yi2018,Terpek}, evasion (adversarial) attacks \cite{Larsson2, Shi1810, Headley2019, Headley2019-1, Silvija2019, Silvija2019-2, Kim1, Kim2, Kim4, Kim5, Kim6, Mao2021, Larsson3, Larsson4, Restuccia2020} and their extensions to secure and covert communications against eavesdroppers \cite{Gunduz1, Gunduz2, Kim3}, causative (poisoning) attacks \cite{Sagduyu1, Luo1, Luo2}, membership inference attacks \cite{MIA, MIA2}, Trojan attacks \cite{Davaslioglu1}, and spoofing attacks \cite{WiseMLSpoofing, YiSpoofing, Cabric} that have been launched against various spectrum sensors and wireless signal (such as modulation) classifiers. Adversarial machine learning has also been considered for NextG by studying evasion and spoofing attacks on deep neural networks (without reinforcement learning) used for NextG spectrum sharing and NextG signal authentication \cite{5Gbookchapter}. In addition, flooding attacks have been considered for NextG network slicing with reinforcement learning \cite{Flooding}.

In this paper, a jamming attack built upon adversarial machine learning is launched against the RL agent that performs resource allocation for NextG network slicing, and the attack exploits the unique properties that (i) the RL algorithm is affected by manipulated rewards and (ii) it takes a while for the RL algorithm to recover even after the attack stops. 

\begin{figure}
	\centering
	\includegraphics[width=0.8\columnwidth]{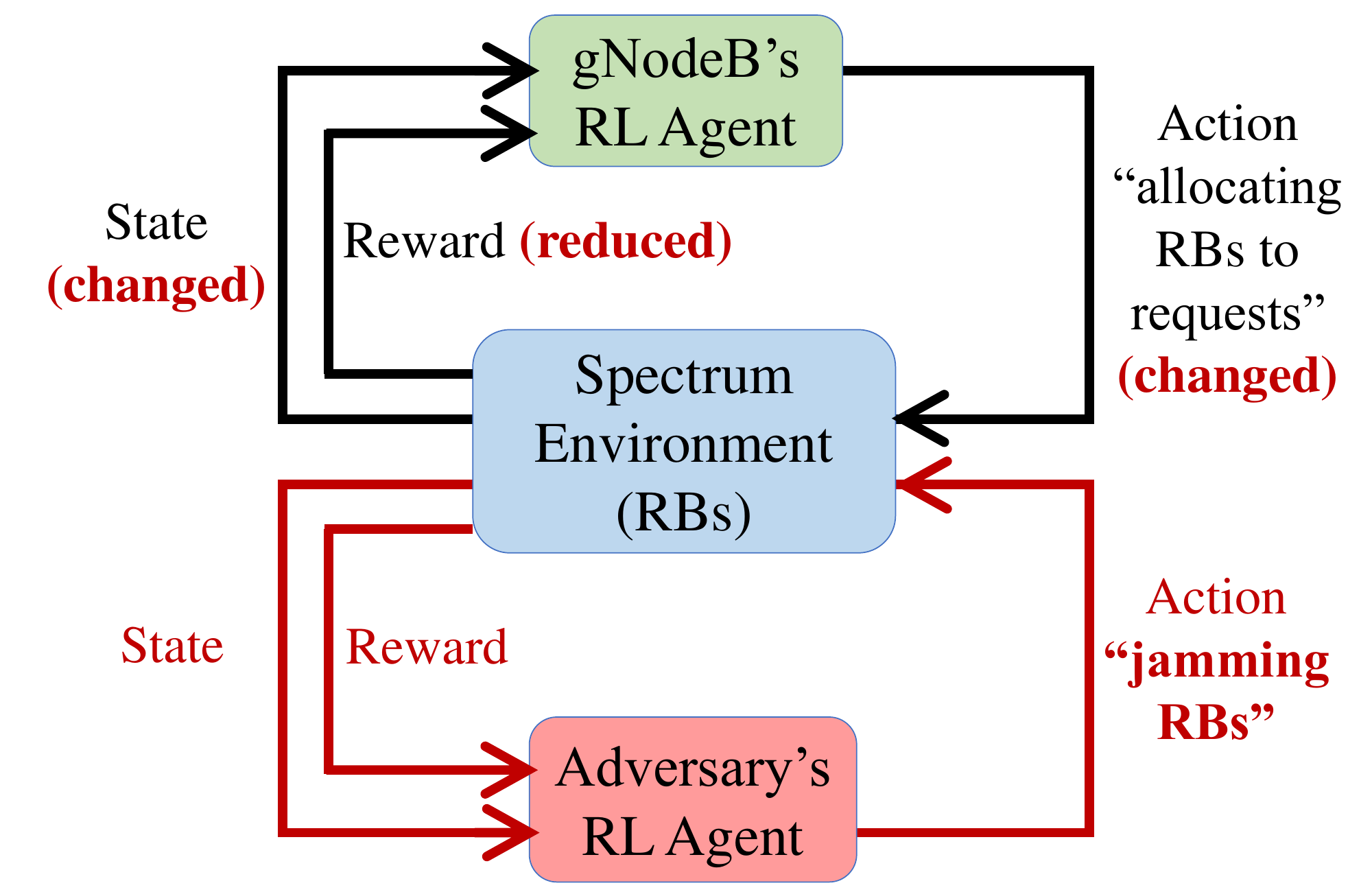}
	\caption{Illustration of how jamming built upon adversarial machine learning manipulates the RL process of the gNodeB to allocate the RBs to network slicing requests.}
	\label{fig:RLattack}
\end{figure}

The \emph{states} of the \emph{surrogate} RL model built by the adversary correspond to the availability of RBs, which are determined by passively sensing the RBs (since the adversary does not have access to the victim RL model, namely it launches a \emph{black-box attack}, and cannot query it with inputs). The \emph{actions} of the adversary are the set of selected RBs to be jammed. We assume that the UEs send a negative acknowledgment (NACK) to confirm a failed transmission from the gNodeB (so that it can be retransmitted later subject to its deadline for reliable communications) and the adversary needs to detect the presence of this feedback without decoding it, as shown in Fig.~\ref{fig:system}.
Typically, the NACK message has a particular pattern: it has a short packet length and it follows data transmission after a fixed time lag. Therefore, it is not difficult to detect the presence of NACK transmissions. The \emph{reward} of the adversary's RL algorithm is the number of jammed and therefore failed requests. 
The RL algorithm at the adversary can learn the effect of its attack and update its RL model (in our example, the Q-table). Once the RL model is well trained, the adversary can make the optimal decision on selecting which RBs to jam by maximizing its expected jamming reward. 
Note that in this attack scenario, the adversary launches an \emph{over-the-air attack} and indirectly manipulates the reward of the RL algorithm by jamming the RBs, as shown in Fig.~\ref{fig:system}.
The interactions between the victim RL algorithm of the gNodeB and the surrogate RL algorithm of the adversary are illustrated in Fig.~\ref{fig:RLattack}.

\begin{figure}
	\centering
	\includegraphics[width=\columnwidth]{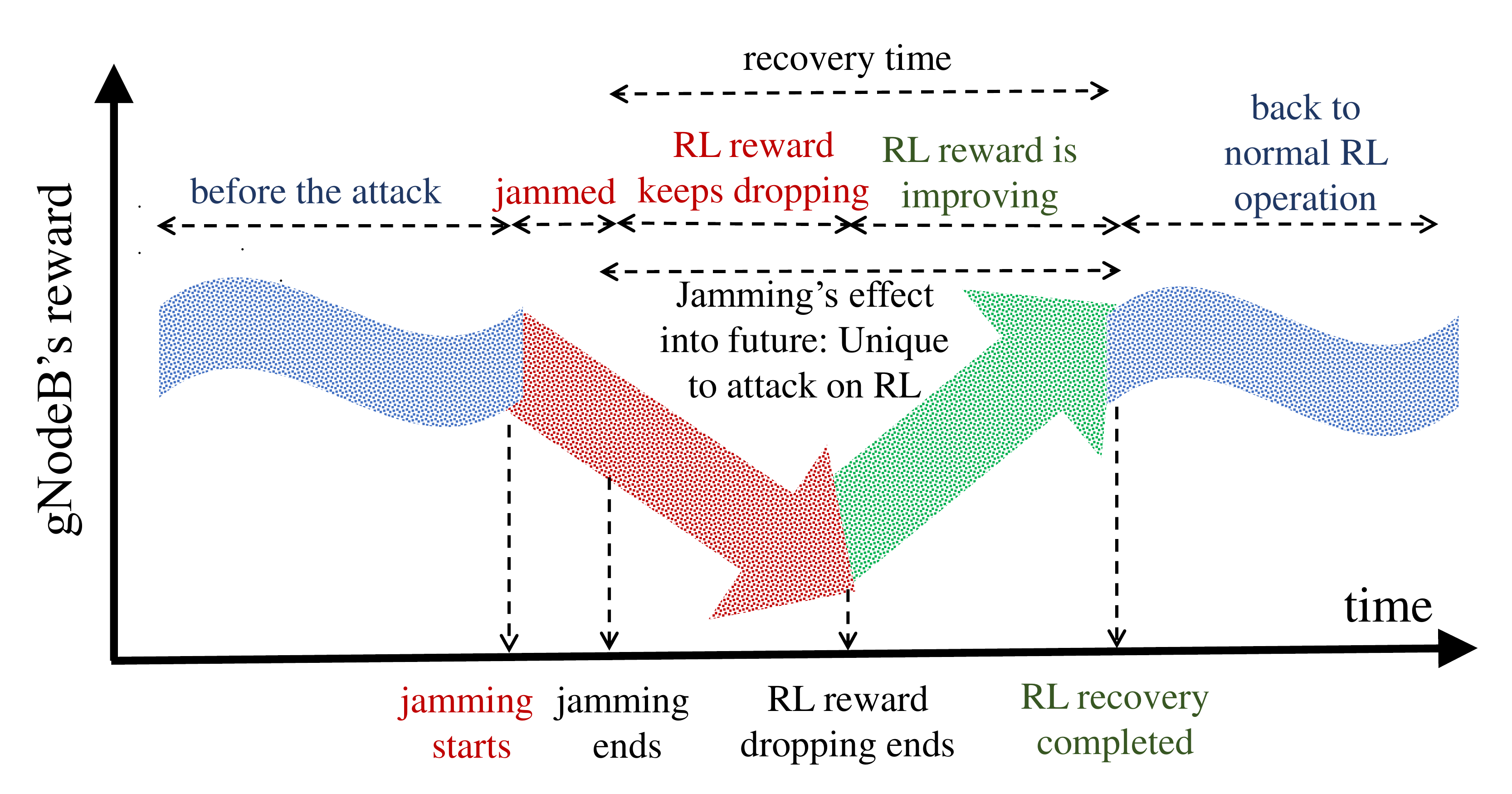}
	\caption{Adversarial machine learning for manipulating the RL process of the gNodeB when it allocates the RBs to network slicing requests.}
	\label{fig:RLperf}
\end{figure}

In performance evaluations, we compare the RL based attack with the \emph{myopic} attack 
and \emph{random jamming} (namely, jamming randomly selected RBs) subject to the same jamming budget constraint. We show that the RL based attack can achieve the largest reduction in the reward of the gNodeB's RL algorithm (under attack and after attack) and the longest recovery time from the attack (after the jamming attack stops). This result demonstrates the adversarial machine learning benefits of manipulating the RL process over a time horizon. As illustrated in Fig.~\ref{fig:RLperf}, \emph{the extension of the attack's impact beyond the time instant when the attack stops is a key capability of the RL based jamming attacks compared to conventional jamming attacks (on data transmissions) whose impact is typically limited to the duration of the attack} (see \cite{trappejamming,sagduyujamming} for examples on conventional jamming attacks on wireless communications).

\subsection{Defense against Adversarial Machine Learning based Attack on NextG Radio Access Network Slicing with Reinforcement Learning}

\begin{figure}
	\centering
	\includegraphics[width=0.8\columnwidth]{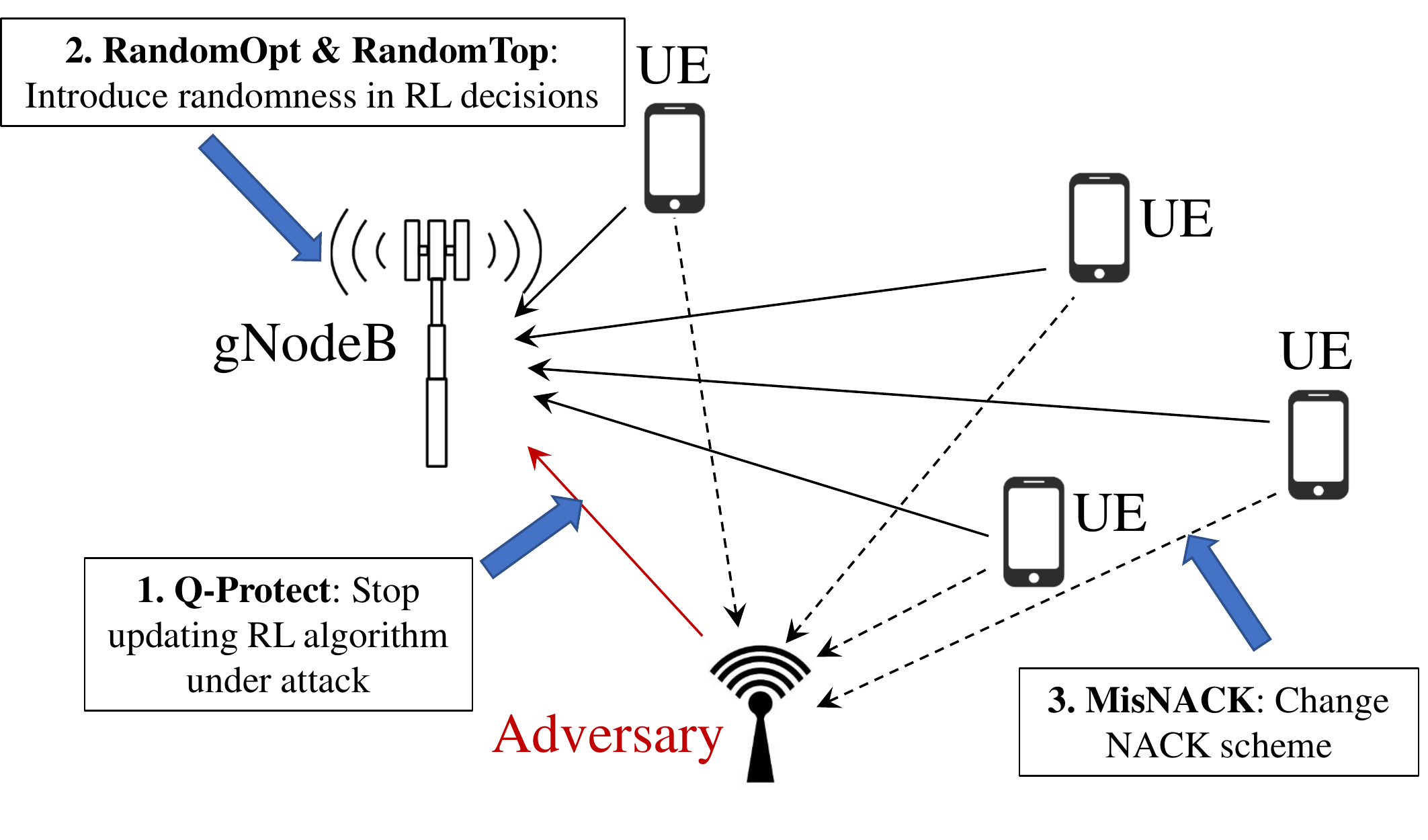}
	\caption{The defense schemes against RL based attack on NextG RAN slicing.}
	\label{fig:defense}
\end{figure}

In this paper, we also investigate  how to \emph{defend} the network slicing operations against the RL based jamming attacks. For that purpose, we introduce three different defense schemes, Q-Protect, RandomOpt/RandomTop, and MisNACK, for the gNodeB or the UE to take (illustrated in Fig.~\ref{fig:defense}):
\begin{enumerate}
\item Q-Protect protects the RL algorithm itself by suspending
the RL algorithm (i.e., Q-table) update once an attack is detected to avoid the impact of the attack on the RL algorithm;
\item RandomOpt and RandomTop
introduce randomness to the decision process in RL (in particular, add perturbations in the Q-table updates) to mislead the learning process of the adversary; or
\item MisNACK manipulates
the feedback (NACK) mechanism such that the adversary may not obtain reliable information to build its attack strategy.
\end{enumerate}
We show that the second defense schemes is more effective than others and can be combined with others to help network slicing operations sustain its performance relative to the case without an attack.

\subsection{Contributions and Paper Organization}

The contributions of this paper are summarized as follows.
\begin{itemize}
\item For an RL based NextG RAN slicing algorithm, we design a novel RL based attack scheme built upon adversarial machine learning that selectively jams the available RBs so that the RL algorithm for network slicing receives incorrect reward (feedback) and updates itself in a wrong way, thereby leading to a significant performance loss of resource allocation for NextG RAN slicing.

\item We design novel defense schemes by considering various characteristics of RL algorithms. The Q-Protect scheme stops the RL algorithm if the reward is unexpected. The RandomOpt and RandomTop schemes make it more challenging for an adversary to learn. The MisNACK scheme provides incorrect information to the adversary.

\item We show the effectiveness of the designed attack and defense schemes using different benchmarks in numerical results. Our results show that the RL based attack scheme achieves better attack performance than benchmark attack schemes, and a combined scheme with multiple defense schemes achieves the best protection.
\end{itemize}

The rest of the paper is organized as follows. Section~\ref{sec:sec2} describes resource allocation for network slicing via RL. Section~\ref{sec:attack} presents the RL based jamming attack that aims to maximize the impact on the gNodeB's performance under the attack and after the attack. Section~\ref{sec:defense} introduces defense schemes to protect the network slicing operations from RL based jamming attacks. Section~\ref{sec:eval} evaluates the attack and defense performances.  Section~\ref{sec:conc} concludes this paper.

\section{The Victim System to Attack: Reinforcement Learning based Resource Allocation for Network Slicing}
\label{sec:sec2}

\begin{figure}
	\centering
	\includegraphics[width=\columnwidth]{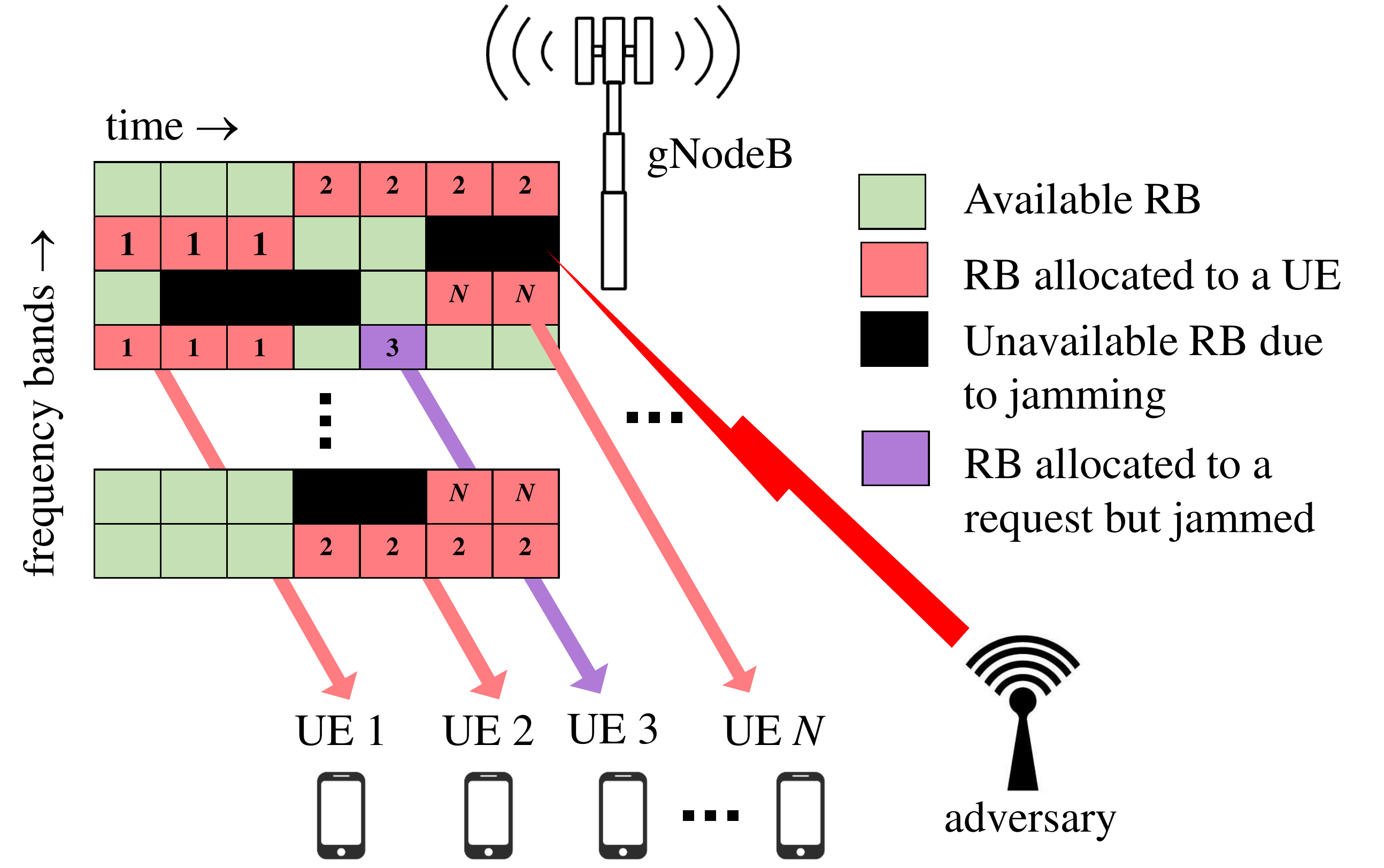}
	\caption{System model for NextG network slicing in the presence of an adversary.}
	\label{fig:sys}
\end{figure}

In this section, we summarize the NextG RAN slicing setting that an adversary aims to attack. We follow the RL formulation of \cite{CAMAD} for network slicing as an example,
while the attack and defense schemes that we consider in the next two sections apply to other RL based NextG RAN slicing settings (e.g., \cite{LiSlicingRL, Koo19:DRL}),
as well. As depicted in Fig.~\ref{fig:sys}, we consider a general scenario in which multiple NextG UEs send requests over time with different QoE requirements, i.e., rate, latency (deadline) and lifetime demands and priority weights, and the gNodeB needs to allocate the RBs to selected requests such that the total weight of served requests over a time period can be maximized. If a request is not granted, it will be kept in a waiting list 
until its deadline expires. There is also an adversary that we will describe in Section~\ref{sec:attack}.

At time slot $t$,
there are a set of active requests $A(t)$ (requests that have just arrived or are in the waiting list). UE $i$'s QoE requirement of rate for its request $j$ is given by
\begin{equation} \label{eq:D}
D_{ij} \geq d_{ij}, \quad (i,j) \in A(t),	
\end{equation}	
where $D_{ij}$ is the achieved downlink data rate and $d_{ij}$ is the minimum required rate. $D_{ij}$ is determined by the assigned bandwidth $F_{ij}$
in an  RB
and the modulation/coding scheme used for communications between the gNodeB and UE $i$.
The data rate (bps) is approximated as \cite{5GNRStd1}:
\begin{equation} \label{eq:datarate}
D_{ij} = c \;K_{ij} \; (1-\textit{BER}_{ij}),
\end{equation}
where $K_{ij}$ 
is the number of aggregated component carriers 
in a band combination
and $\textit{BER}_{ij}$ is the bit error rate 
of UE $i$ for its request $j$ (which depends on the signal-to-noise ratio (SNR) and is computed for AWGN channel with low-density parity-check 
coding), and constant $c$ is approximately $12.59 \times 10^6$ when a single-antenna UE uses QPSK modulation, $60$ kHz subcarrier spacing and $10$ MHz bandwidth.

The constraints of resource assignments to network slices are given by
\begin{eqnarray}
\sum_{i,j} F_{ij} \; x_{ij}(t) \leq  F(t), \:\: (i,j) \in A(t), \label{eq:assign-f} 		
\end{eqnarray}
where $F_{ij}$ is the assigned bandwidth and $F(t)$ represents the available communication resources (RBs) of the gNodeB at time $t$ (resources that are assigned previously to some requests and not terminated yet become temporarily unavailable) and $x_{ij} (t)$ is the binary indicator on whether UE $i$'s request $j$ is satisfied at time $t$.

By considering the optimization problem for a time horizon, the resources are updated from time $t-1$ to time $t$ as
\begin{eqnarray}
F(t) & = & F(t-1)+F_r (t-1)-F_a (t-1), \label{eq:update-f}
\end{eqnarray}			
where $F_r (t-1)$ and $F_a (t-1)$ are the released and allocated resources on frequency at time $t-1$, respectively. 
Each request has a lifetime $l_{ij}$ and if it is satisfied at time slot $t$ (namely, the service starts in time slot $t$), this request will end at the end of time slot $t+l_{ij}-1$.
The released and allocated resources at time $t$ are given by
\begin{eqnarray}
F_r (t) & = & \sum_{(i,j) \in R(t)} F_{ij},	\label{eq:release-f} \\
F_a (t) & = & \sum_{i,j} F_{ij}, \label{eq:allocate-f}				
\end{eqnarray}				
where $R(t)$ denotes the set of requests ending (completed or expired) at time $t$.
Then, the optimization problem is given by
\begin{eqnarray}
\max_{x_{ij}(t)} \sum_t \sum_{ij}  w_{ij} \; x_{ij}(t), \:\: (i,j) \in A(t)
\label{eq:opt2}
\end{eqnarray}
subject to (\ref{eq:D})--(\ref{eq:allocate-f}), where $w_{ij}$ is the weight for UE $i$'s request $j$ to reflect its priority.

As a model-free RL algorithm, we use Q-learning to learn the policy that determines which action (resource assignment) to take under a given state (available resources and requests) for the gNodeB. The gNodeB applies Q-learning to compute the function $Q:S\times A\to \mathbb {R}$ (maintained as the Q-table) to evaluate the quality of action $A$ producing reward $R$ at state $S$. At each time $t$, the gNodeB selects an action $a_{t}$, observes a reward $r_{t}$, and transitions from the current state $s_t$ to a new state $s_{t+1}$ (this transition depends on current state $s_{t}$ and action $a_t$), and updates $Q$.

Initializing $Q$ as a random matrix and using the weighted average of the old value and the new information, Q-learning performs the value iteration update for $Q$ as follows:
\begin{eqnarray}
\label{eq:qfunction}
Q(s_{t},a_{t}) \leftarrow && \hspace{-0.6cm} Q(s_{t},a_{t}) \\ && \hspace{-0.6cm} +  {\alpha } \cdot \left( {r_{t}} + {\gamma } \cdot \max_{a} Q(s_{t+1},a) - {Q(s_{t},a_{t})} \right), \nonumber
\end{eqnarray}
where $\alpha$ is the learning rate ($0<\alpha \leq 1$) and $\gamma$ is the discount factor ($0\leq \gamma \leq 1$) for rewards over time. As the size of the states increases, it becomes computationally more efficient to approximate the Q-function by training a deep neural network, leading to a deep Q-network formulation.

In dynamic resource allocation to network slices, the reward at time $t$ is $w_{ij}$ if UE $i$'s request $j$ is satisfied at time $t$, i.e., $x_{ij}(t) = 1$. Note that the reward measures the satisfied QoE demands of network slices and therefore it indirectly reflects the achieved QoE performance such as throughput and delay.

An action is to assign resources to a request at time $t$. Multiple actions can be taken at the same time instance.
The states at $t$ are $F$ binary variables on the availability of $F$ RBs and $(F_{ij}, w_{ij})$ for a request under consideration. The state transition at time $t$ is driven by allocating resources for requests granted at time $t$ and releasing resources after lifetimes of some active services expire at time $t$. In particular, the state transitions are given by (\ref{eq:update-f})-(\ref{eq:allocate-f}).
The states, actions, and rewards of the RL algorithm for network slicing are summarized in Table~\ref{table:RL}.
The standard Q-learning algorithm of (\ref{eq:qfunction}) is considered.
The notation used in this paper is shown in Table~\ref{table:notation}.

\begin{table}
	\caption{RL algorithm for network slicing.}
	\centering
	{\vspace{0.2cm}
	\footnotesize
		\begin{tabular}{c|c}
RL term & Specification (at any given time instant) \\
\hline \hline
State & Availability of RBs, active requests \\ \hline
Action & Assign RBs to selected network slicing requests \\ \hline
Reward & Total weights of satisfied network slicing requests
		\end{tabular}
	}
	\label{table:RL}
\end{table}

\begin{table}
	\caption{Notation table.}
	\centering
	{\vspace{0.2cm}
	\footnotesize
		\begin{tabular}{c|l}
Symbol & Definition \\
\hline \hline
$a_{t}$ & Action at time slot $t$ \\ \hline
$A(t)$ & Set of active requests at time $t$ \\ \hline
$B$ & Maximum number of RBs that the adversary can jam at any \\
& given time \\ \hline
$\textit{BER}_{ij}$ & Bit error rate for UE $i$'s request $j$ \\ \hline
$d_{ij}$ & Minimum required rate for UE $i$'s request $j$ \\ \hline
$D_{ij}$ & Achieved downlink data rate for UE $i$'s request $j$ \\ \hline
$F$ & Number of all RBs \\ \hline
$F_{ij}$ & Assigned RBs for UE $i$'s request $j$ \\ \hline
$F(t)$ & Available RBs of the gNodeB at time $t$ \\ \hline
$F_r (t)$ & Released RBs on frequency at time $t$ \\ \hline
$F_a (t)$ & Allocated RBs on frequency at time $t$ \\ \hline
$l_{ij}$ & Lifetime of UE $i$'s request $j$ \\ \hline
$r(t)$ & Reward at time $t$ \\ \hline
$r_{\text{top}}$ & Percentage to determine whether a reward is considered as \\
& top reward or not \\ \hline
$R(t)$ & Set of requests ending at time $t$ \\ \hline
$s(t)$ & State at time $t$ \\ \hline
$w_{ij}$ & Weight for UE $i$'s request $j$ to reflect its priority \\ \hline
$x_{ij} (t)$ & Binary indicator on whether UE $i$'s request $j$ is satisfied \\
& at time $t$ \\ \hline
$\alpha$ & Learning rate of the Q-learning algorithm \\ \hline
$\gamma$ & Discount factor for rewards over time
		\end{tabular}
	}
	\label{table:notation}
\end{table}

\section{Attack on Reinforcement Learning for NextG Network Slicing} \label{sec:attack}

We now consider an adversary that attacks
an RL algorithm for NextG RAN slicing, e.g., the one discussed in Section~\ref{sec:sec2}.
Other example victim systems include the RL based network slicing schemes in \cite{LiSlicingRL, Koo19:DRL}.

\subsection{Reinforcement Learning based Attack}
\label{sec:RLAttack}

Since RL keeps collecting data and updating itself, it has \emph{two unique properties} that we leverage to build and evaluate attacks on RL.
\begin{enumerate}
\item If an adversary changes the state or the reward, it can affect the RL algorithm.

\item On the other hand, if the adversary stops attacking, the RL algorithm will recover by itself.
\end{enumerate}
In this section, we exploit the first property to design the attack on the RL algorithm of the NextG network slicing. As this attack can still affect the RL significantly even after the attack stops for a while, we measure the impact due to the second property in Section~\ref{sec:eval}.

To launch an attack, the adversary can change either the state or the reward of the  RL agent. For the RL algorithm presented in Section~\ref{sec:sec2}, the state includes the RB availability and a request under consideration. Both are maintained by the gNodeB. Therefore, they cannot be changed by the wireless adversary that is physically separated from the gNodeB and does not have direct access to the gNodeB's RL algorithm. On the other hand, the adversary can affect the reward if it jams an RB to be allocated to a request.
In that case, the request will not be successful even if resources are allocated by the RL algorithm and there is no reward gained by the RL algorithm.

We assume a practical constraint that the adversary has \emph{limited jamming capability} (typically due to limited energy budget) and thus cannot jam all RBs to maximize its impact. We denote $B$ as the maximum number of RBs that the adversary can jam at any given time. Due to this constraint, it is important for the adversary to select the RBs that are available and likely to be allocated such that jamming these RBs can affect network slicing requests to be selected by the RL algorithm.

The ideal case is that the adversary can build a \emph{surrogate model} (another RL algorithm) that can predict which RBs will be allocated and then use the predicted results to decide which RBs should be jammed.
However, this case is impractical since (i) the request under consideration is a part of the state, which is unknown to the adversary, and (ii) the reward is the request's weight, which is unknown to the adversary. Therefore, the adversary builds a different RL model (as an approximate surrogate model).
Although the RL algorithm is the same as that discussed in Section~\ref{sec:sec2}, this RL model of the adversary has different state, action, and reward properties given as follows.
\begin{itemize}
\item The \emph{state} is the set of binary variables that indicate the availability of all RBs.

\item An \emph{action} corresponds to selecting the set of $\min\{B, F(t)\}$ RBs from $F(t)$ available RBs, and jamming those selected RBs. Note that there is also the action of not jamming any RB. Thus, the number of possible actions is $C_{F(t)}^B +1$ (where $C_{F(t)}^B$ is the number of $B$-combinations from a set of $F(t)$ elements, i.e., the number of possibilities in picking $B$ out of $F(t)$) if $F(t) > B$, or $2$ (jam or not) if $F(t) \le B$.

\item The \emph{reward} is the number of jammed requests at a given time. We assume that there is a NACK transmitted from a NextG UE at the end of a time slot if the
transmission
is not successful. If the adversary jams an RB and later observes the NACK, the reward on this channel is $1$. Note that the adversary does not need to decode the NACK. It needs to detect the presence of NACK only, which is possible by distinguishing the NACK from data transmissions (as the NACK is shorter than data portion and has the structure of appearing between requests and data transmissions).
\end{itemize}
To initialize the Q-table, we set entries in the column of no jamming to zeros and entries in other columns to the number of jammed RBs.

The adversary applies RL to update its Q-table by (\ref{eq:qfunction}) and to take actions based on its Q-table. The states, actions, and rewards of the adversary's RL algorithm are summarized in Table~\ref{table:ARL}.

\begin{table}
	\caption{The adversary's RL algorithm.}
	\centering
	{\vspace{0.2cm}
	\footnotesize
		\begin{tabular}{c|c}
RL term & Specification (at any given time instant) \\
\hline \hline
State & Availability of RBs \\ \hline
Action & Jam selected RBs \\ \hline
Reward & Number of jammed requests
		\end{tabular}
	}
	\label{table:ARL}
\end{table}

\subsection{Performance Metrics and Benchmark Attack Schemes}

When the adversary launches its attack, we can observe the performance reduction of the gNodeB by comparing it with the case of no attack. The reason for the performance loss is that some requests fail due to jamming and thus their weights are not counted in the reward of the gNodeB.

More interestingly, since some rewards are changed by jamming the RBs and the gNodeB's RL algorithm is updated based on these changed rewards, the attack also affects the RL algorithm itself.
As a result, even if the adversary stops jamming the RBs, the performance of NextG network slicing cannot return to previous levels (before the attack) right away.
Instead, it takes some time for the gNodeB to collect sufficient data to correct its algorithm and then finally its performance can go back to the case when there is no attack.
To measure this impact after the attack stops, we consider the following metrics.
\begin{itemize}
\item \textit{Recovery time}: The time it takes (after the attack, namely jamming, stops) for the network slicing performance (namely, the reward) to go back to ``normal" (the level before the attack).
The recovery time is an important metric since if it is long, the adversary can stop its attack to avoid being detected or to save energy and then start its attack again before the recovery time.

\item \textit{Maximum performance reduction}: The maximum gap in performance compared to the normal (before-the-attack) value during the recovery time. The performance is measured as the running averaged reward.
The maximum performance reduction describes the maximum impact during the recovery time.

\item \textit{Total performance reduction}: The accumulated performance gap to the normal value during the recovery time.
The total performance reduction is a more robust metric than the above two, since it is not affected by small performance reduction (comparing with recovery time) or single extreme point (comparing with maximum performance reduction).
\end{itemize}

In addition to this attack, we also consider the case of no attack and two benchmark attacks, namely random attack and myopic attack, for performance evaluation:
\begin{itemize}
	\item \textit{Random attack}: The adversary randomly jams some RBs (that are uniformly randomly selected from all RBs) subject to the jamming budget.
	\item \textit{Myopic attack}: The adversary selects which RBs to jam (subject to the jamming budget) with the objective of maximizing the instantaneous reward without the consideration of future rewards.
\end{itemize}

Note that the proposed RL based attack takes time to improve its attack actions as its RL algorithm learns how to attack NextG RAN slicing. Other attacks schemes do not have this of process of gradual improvement. As we measure the recovery time, maximum and total performance reduction over the same period of time (including the warm-up time) for all attack schemes, we provide a fair comparison of RL based attacks with random and myopic attack.
The performance of these attacks is evaluated in Section~\ref{sec:eval}.

\section{Defense against Attacks on Reinforcement Learning for NextG Network Slicing}
\label{sec:defense}

To protect the RL based resource allocation for NextG RAN slicing (e.g., \cite{CAMAD, LiSlicingRL, Koo19:DRL})
from the RL based jamming attacks, we present different defense schemes (illustrated in Fig.~\ref{fig:defense}) for the gNodeB or the UE to take.

\begin{enumerate}
\item \textit{Q-Protect}: One \emph{reactive} defense scheme is based on protecting the RL algorithm itself. Note that if there is no attack, once a network slicing request is served, some reward is expected. However, if the RBs that are allocated to this request are jammed, this request cannot be satisfied and therefore its reward is reduced to zero. Thus, the gNodeB can detect the jamming attack by checking the changes in the reward.
For numerical results, we assume that the
attack
is detected if the running average of the rewards drops by $10\%$. Hence, the gNodeB
suspends
the Q-table update once an attack is detected to avoid the impact of the attack on the RL based network slicing algorithm.
We call this defense scheme ``Q-Protect", which can be applied to any RL algorithm.
The adversary cannot force the gNodeB to update its RL algorithm and thus cannot circumvent this defense.

\item \textit{RandomOpt} and \textit{RandomTop}: A \emph{proactive} defense scheme aims to manipulate the adversary's learning process (namely, its surrogate model). This defense scheme can be effective against any learning-based attack. However, it cannot protect network slicing from random jamming attacks. The gNodeB can proactively introduce randomness to the resource allocation actions in its RL algorithm such that an adversary cannot easily learn how to build its RL algorithm. We propose two defense schemes, with and without performance loss when there is no attack
\begin{enumerate}
\item Note that there may be multiple best actions with the same reward in the Q-table. Then, the gNodeB can randomly select any action without any performance loss.\footnote{To simplify discussion, we assume that the Q-table is perfect and thus the same reward in the Q-table means the same long-term reward in the objective. In reality, the Q-table may not be perfect and thus there can still be performance loss under this policy.}
We call this defense scheme ``RandomOpt", which can be applied to any algorithm that can find multiple optimal actions.
\item The randomness among best actions may not be sufficient to mitigate the performance loss due to the attack. Another defense scheme is to randomly select an action from top actions (those with rewards that are close to the best reward).
An action is considered as ``Top" if its reward is at least $r_{\text{top}}$ percentage of the maximum reward.
This defense scheme introduces more randomness but may incur performance loss even if there is no attack.
We call this defense scheme ``RandomTop", which can be applied to any algorithm that can find multiple near-optimal actions.
The adversary cannot remove the randomness introduced by the defender and thus cannot circumvent these two defense schemes.
\end{enumerate}

\item \textit{MisNACK}: Another \emph{proactive} defense scheme aims to manipulate the feedback (NACK) mechanism such that the adversary may not obtain reliable information to build its attack strategy. We note that the UE sends a NACK over any jammed RB if some of its RBs are jammed. That is, there is one NACK transmitted for each failed request. The adversary monitors the jammed RBs to detect the presence of NACK transmission and thus defines the reward of its action. As a defense, each UE can send the NACK over an unjammed RB (if any) such that no NACK can be detected by the adversary that monitors only the channel that it has jammed. If all its RBs are jammed, the UE can send multiple NACKs over these RBs such that the adversary will overestimate the effect of its attack. This way, the adversary reduces the reliability of NACK for the adversary.
We call this defense scheme ``MisNACK", which can be applied to any algorithm that uses NACK.
The adversary cannot force UEs not to send misleading NACKs and thus cannot circumvent this defense.
\end{enumerate}

The performance of these defense schemes is evaluated in Section~\ref{sec:eval}.

\section{Performance Evaluation} \label{sec:eval}

Suppose that the gNodeB receives requests from $30$ UEs. For each UE, requests arrive with the rate of $0.05$ per slot. Here, a slot corresponds to each time block which is $0.23$ ms long with $60$ kHz subcarrier spacing. For each request, the weight of a request is assigned (uniformly) randomly in $[1,5]$, the lifetime is assigned randomly in $[1,10]$ slots, and the deadline is assigned randomly in $[1,20]$ slots. The maximum received SNR is selected randomly from [1.5,3]. The total frequency is $10$ MHz and is split into $11$ bands, i.e., there are $11$ RBs. We also consider a scenario with a smaller number of RBs, namely $5$ RBs.

\subsection{Attack Performance Evaluation}
The same scenario over $1000$ time slots is repeated to evaluate these attacks.
For Q-learning, we set the discount factor as $\gamma=0.95$ and the learning rate as $\alpha=0.1$.

\begin{table}
	\caption{Performance comparison of Q-learning and other attacks when there are $11$ RBs.}
	\centering
	{\vspace{0.2cm}
	\footnotesize
		\begin{tabular}{c|c|c|c|c}
Attack & Maximum & Recovery & Maximum & Total \\
scheme & jammed & time & reduction & reduction \\
 & RBs & & in reward & in reward \\
\hline \hline
 & 1 & 1038 & 1.447 & 736.216 \\ \cline{2-5}
 & 2 & 1191 & 1.801 & 911.604 \\ \cline{2-5}
Q-learning & 3 & 1548 & 1.957 & 1006.174 \\ \cline{2-5}
 & 4 & 2086 & 2.014 & 1038.988 \\ \cline{2-5}
 & 5 & 2038 & 2.714 & 1410.069 \\ \hline
 & 1 & 1035 & 1.343 & 670.071 \\ \cline{2-5}
 & 2 & 1060 & 1.587 & 788.289 \\ \cline{2-5}
Myopic & 3 & 1028 & 1.684 & 836.998 \\ \cline{2-5}
 & 4 & 1207 & 1.775 & 894.721 \\ \cline{2-5}
 & 5 & 1365 & 1.772 & 889.113 \\ \hline
 & 1 & 1197 & 1.000 & 506.947 \\ \cline{2-5}
 & 2 & 1233 & 1.489 & 750.976 \\ \cline{2-5}
Random & 3 & 1170 & 1.813 & 907.546 \\ \cline{2-5}
 & 4 & 1180 & 2.061 & 1032.088 \\ \cline{2-5}
 & 5 & 1202 & 2.273 & 1141.359
		\end{tabular}
	}
	\label{table:1}
\end{table}

\begin{figure}
	\centering
	\includegraphics[width=0.85\columnwidth]{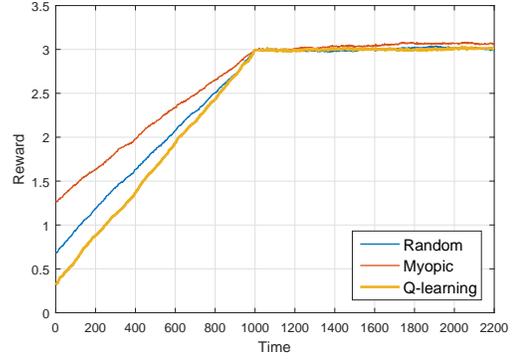}
	\caption{The reward of RL algorithm for NextG RAN slicing after the attack stops when there are $11$ RBs.}
	\label{fig:after11}
\end{figure}

We assume that the adversary launches its attack over $10000$ slots. The benchmark of no attack case is also run over $10000$ slots in total and the achieved reward is measured as $3.032$ over the first $1000$ slots (and this is used as the benchmark for recovery). Then, we measure the average reward over the past $1000$ slots after the attack stops and once this average reward reaches $3.032$, namely when the system performance is assumed to recover from the attack. We also measure the performance gap to the benchmark and present results on the maximum gap and the total gap during the recovery time.

\begin{figure}
	\centering
	\includegraphics[width=0.85\columnwidth]{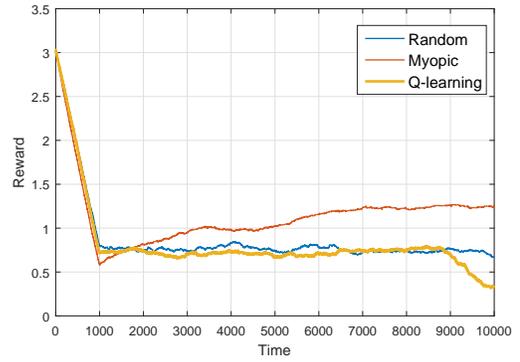}
	\caption{The reward of RL algorithm for NextG RAN slicing under the attack when there are $11$ RBs.}
	\label{fig:under11}
\end{figure}

\begin{table}
	\caption{Performance comparison of RL based and other attacks when there are $5$ RBs.}
	\centering
	{\vspace{0.2cm}
	\footnotesize
		\begin{tabular}{c|c|c|c|c}
Attack & Maximum & Recovery & Maximum & Total \\
algorithm & jammed & time & reduction & reduction \\
 & RBs & & in reward & \\
\hline \hline
Q-learning & 1 & 990 & 0.476 & 253.221 \\ \cline{2-5}
 & 2 & 1100 & 0.799 & 418.045 \\ \hline
Myopic & 1 & 925 & 0.370 & 175.576 \\ \cline{2-5}
 & 2 & 992 & 0.583 & 283.019 \\ \hline
Random & 1 & 1029 & 0.403 & 199.924 \\ \cline{2-5}
 & 2 & 1003 & 0.620 & 308.266 \\
		\end{tabular}
	}
	\label{table:2}
\end{table}

For comparison purposes, we obtain results for attacks by random and myopic jamming attacks in Table~\ref{table:1}, where results for random jamming are averaged over $20$ runs. The RL based attack (introduced in Section \ref{sec:RLAttack}) has longer and larger impact on the NextG network slicing performance than other attacks,
which means that the RL based attack has better performance. Depending on the maximum number of jammed RBs, Q-learned based attack increases the recovery time by up to $77\%$, increases the maximum reduction in reward up by to $53\%$, and increases the total reduction in reward by up to $59\%$ compared to benchmark attack schemes.
We show in Fig.~\ref{fig:after11} how the reward changes over time after the attack stops
when the maximum number of jammed RBs is $5$.
The advantage of RL based attack comes from the smallest reward when the attack stops. Thus, we also check the reward under different attacks.
The RL algorithm's performance under different attacks is shown in Fig.~\ref{fig:under11}. Since we show the average reward over the past $1000$ slots, the performance is high at the beginning and decreases fast. Then, the performance under random jamming or Q-learning based jamming remains still small while the performance under myopic jamming keeps increasing. This is because the myopic algorithm is deterministic and thus it is easy to learn and mitigate it by the RL algorithm for NextG RAN slicing.

\begin{figure}
	\centering
	\includegraphics[width=0.85\columnwidth]{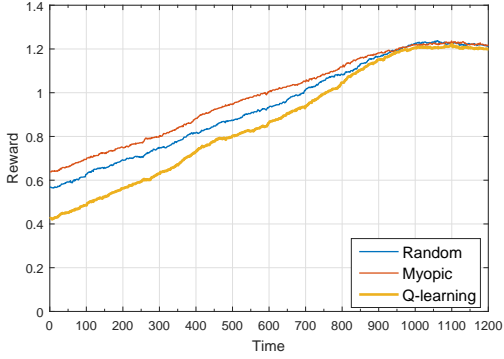}
	\caption{The reward of RL algorithm for NextG RAN slicing after the attack stops when there are $5$ RBs.}
	\label{fig:after5}
\end{figure}

\begin{figure}
	\centering
	\includegraphics[width=0.85\columnwidth]{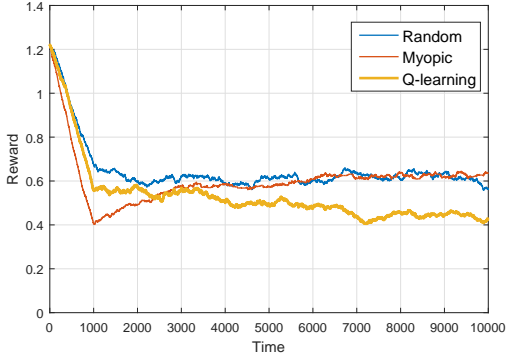}
	\caption{The reward of RL algorithm for NextG RAN slicing under the attack when there are $5$ RBs.}
	\label{fig:under5}
\end{figure}

Next, we evaluate the performance when the number of RBs is reduced from $11$ to $5$ (other parameters remain the same). Results are shown in Table~\ref{table:2}. As before, the Q-learning  based attack has longer and larger impact on the performance than benchmark attacks. Fig.~\ref{fig:after5} and Fig.~\ref{fig:under5} show the reward over time after the attack stops and under the attack, respectively. The trends in Fig.~\ref{fig:after5} and Fig.~\ref{fig:under5} are the same as the trends observed in Fig.~\ref{fig:after11} and Fig.~\ref{fig:under11} when there are $11$ RBs.
On the other hand, we may increase the number of RBs. We find that the Q-table size will be increased by the second order of the number of RBs. A large Q-table requires both long training time and large memory usage. It would be better to design a solution using deep Q-learning instead.

\subsection{Defense Performance Evaluation}

\begin{figure}
	\centering
	\includegraphics[width=0.85\columnwidth]{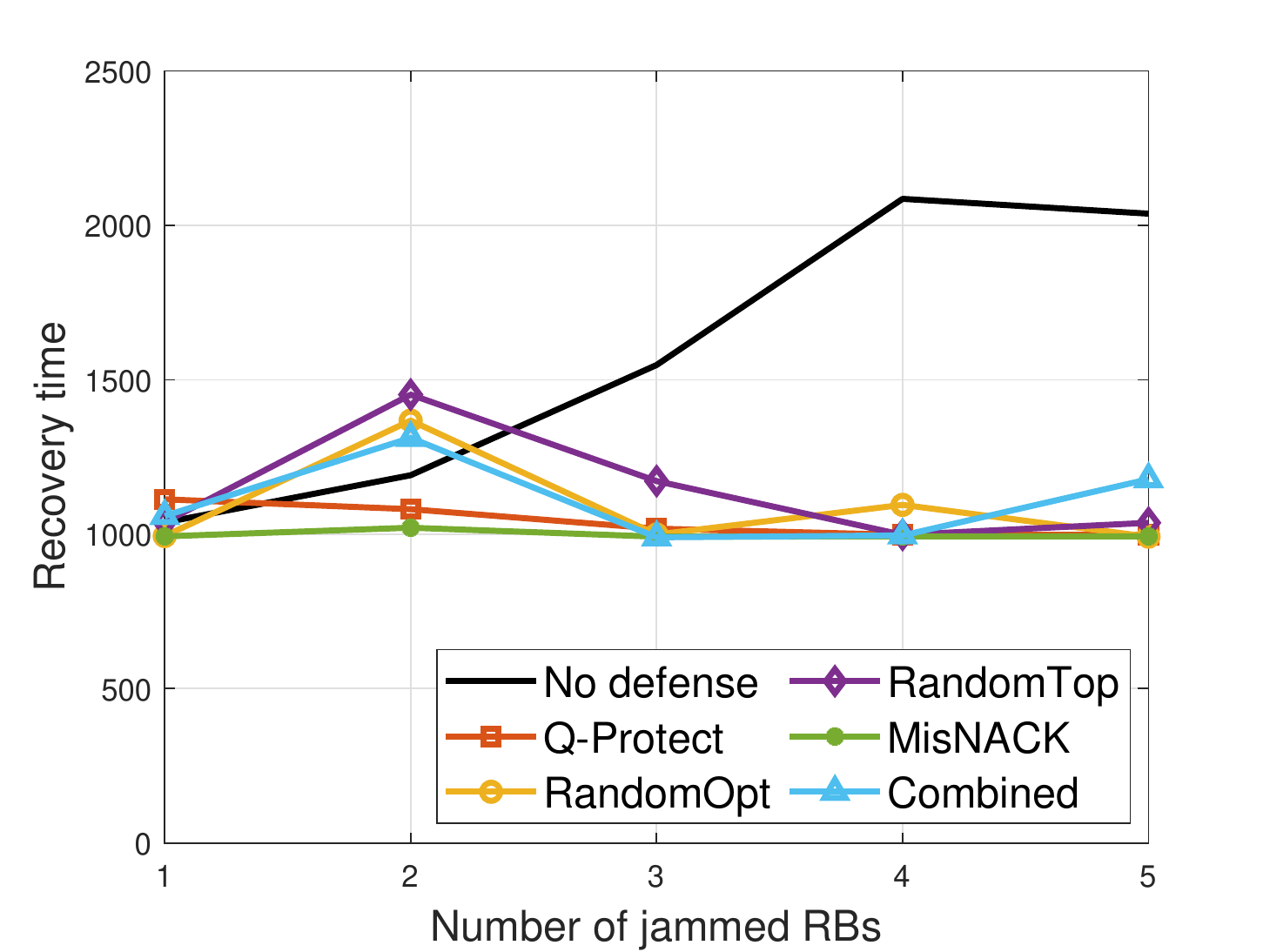}
	\caption{Recovery time of  RL algorithm for NextG RAN slicing under different attacks.}
	\label{fig:recovery}
\end{figure}

We now present the performance of different defense schemes (described in Section~\ref{sec:defense}) against the RL based attack.
For RandomTop, an action is considered as ``Top" if its reward is at least $r_{\text{top}}=50$\% of the maximum reward.
The recovery time for the no defense case and all defense schemes is shown in Fig.~\ref{fig:recovery}, where the ``Combined" scheme combines different defense schemes (``Q-Protect", ``RandomTop", and ``MisNACK") and apply them jointly to strengthen the overall defense against the RL based attack on NextG network slicing.
In particular, ``Q-Protect" aims to protect the Q-table while both ``RandomTop" and ``MisNACK" aim to attack the adversary's learning process, and thus they all can be combined.
Compared with the no attack case, all the defense schemes reduce the recovery time if the number of jammed RBs is at least three. The improvement when there is one jammed RB is not significant. The random effect in ``RandomOpt" and ``RandomTop" makes them worse than the no defense case if the number of jammed RBs is $2$. In fact, although it takes long time to recover, the amount of reduction in reward is not large. Thus, we further study the reduction in reward, in terms of the maximum reduction and the total reduction, during the recovery period.

\begin{figure}
	\centering
	\includegraphics[width=0.85\columnwidth]{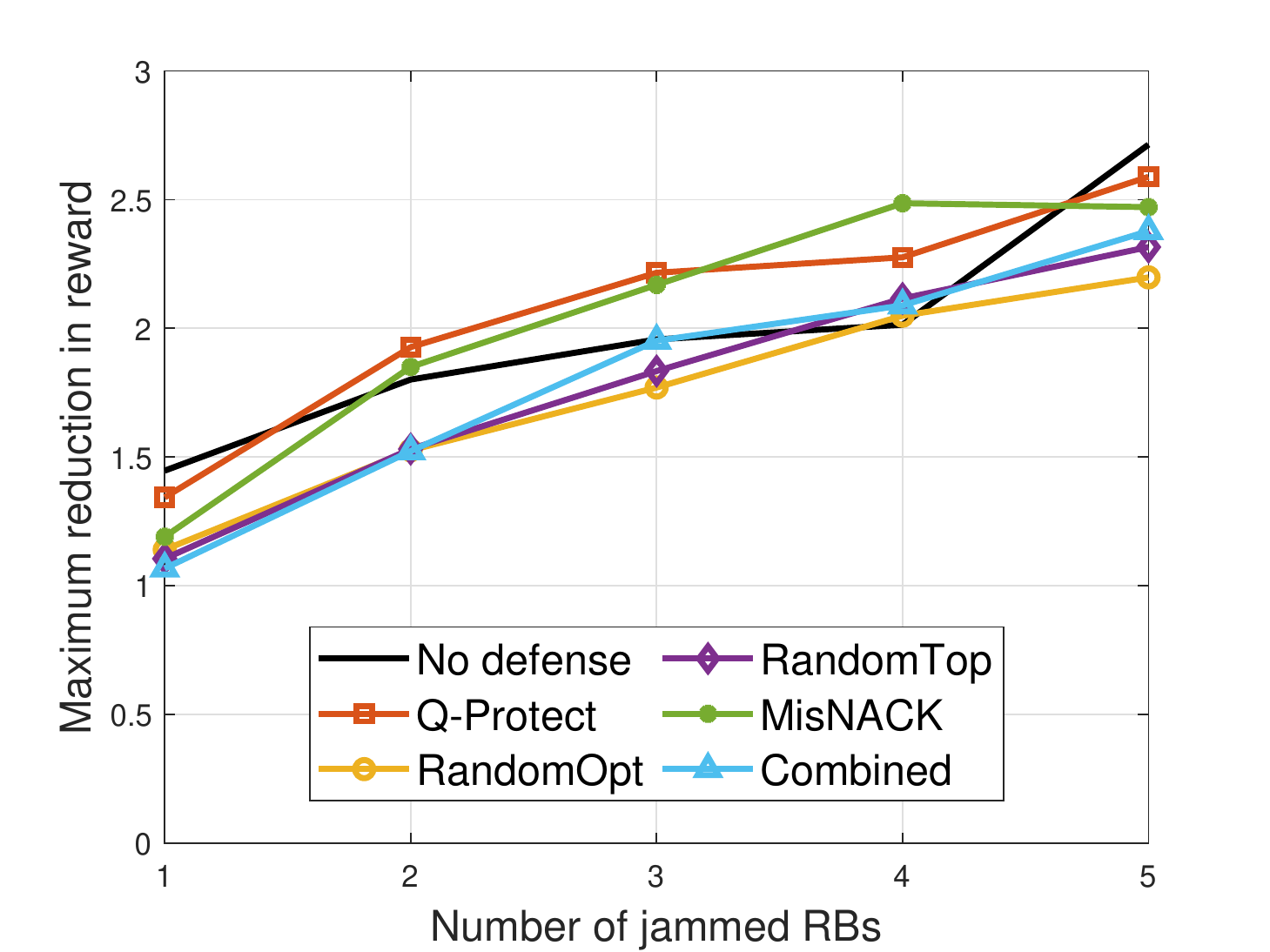}
	\caption{Maximum reduction in the reward of RL algorithm for NextG RAN slicing under different attacks.}
	\label{fig:maxreduction}
\end{figure}

The maximum reduction in reward for the no defense case and all defense schemes are shown in Fig.~\ref{fig:maxreduction}. Compared with the no attack case, the ``RandomOpt", ``RandomTop", and ``Combined" schemes achieve smaller reduction in most of the cases.



\begin{figure}
	\centering
	\includegraphics[width=0.85\columnwidth]{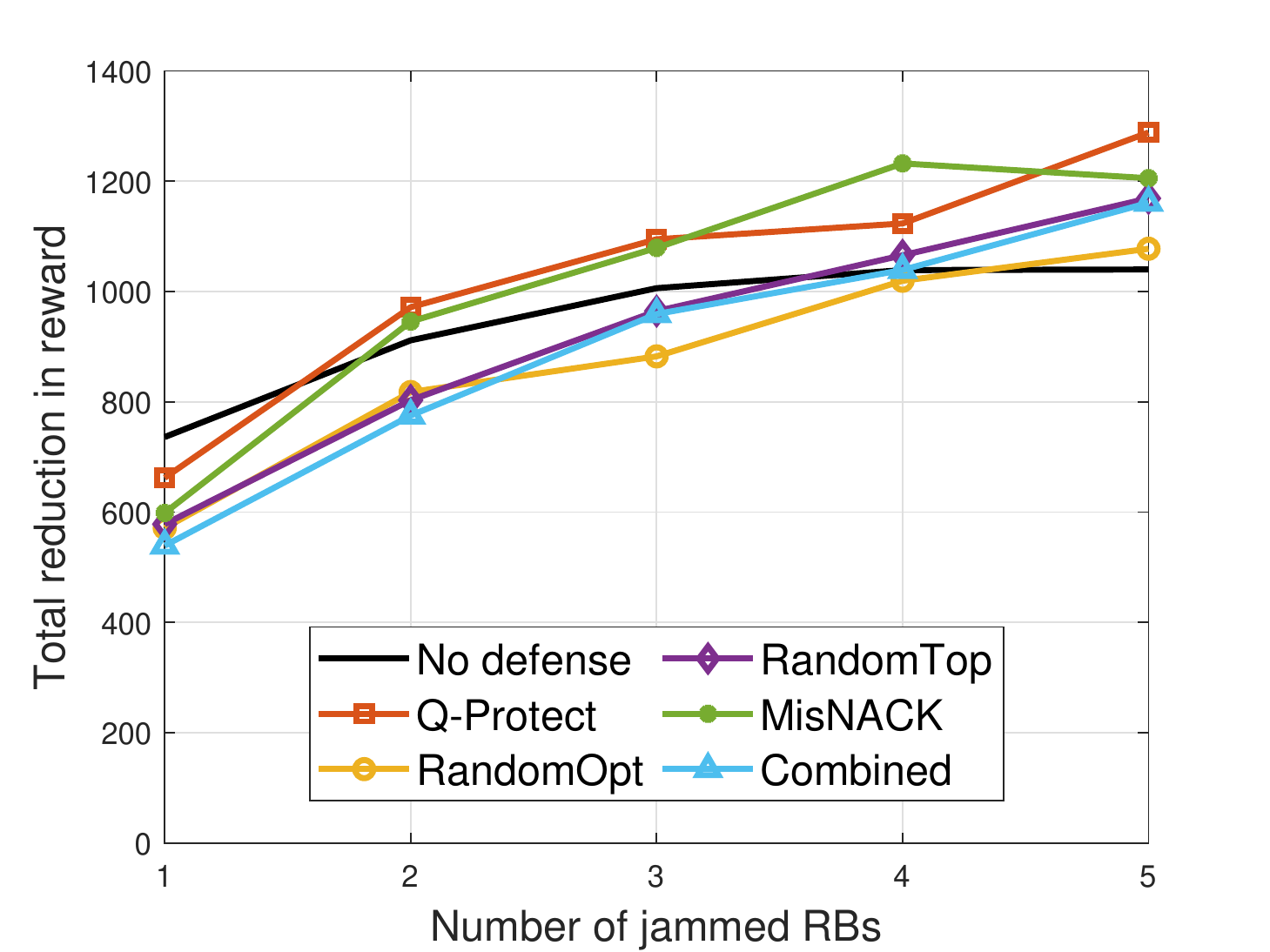}
	\caption{Total reduction in the reward of RL algorithm for NextG RAN slicing under different attacks.}
	\label{fig:totreduction}
\end{figure}

The total reduction in reward for the no defense case and defense schemes is shown in Fig.~\ref{fig:totreduction}. Compared with the no attack case, the ``RandomOpt", ``RandomTop", and ``Combined" schemes achieve smaller reduction for most cases.

In summary, the ``Combined" scheme achieves better defense performance than other defense schemes for most of cases. Therefore, we evaluate the performance of the ``Combined" scheme in further detail. The performance when there are $5$ RBs is shown in Table~\ref{table:randomtop5}. Compared with the results in in Table~\ref{table:2}, we note that the ``Combined" scheme can improve the performance for NextG RAN slicing, and this observation holds for both cases with $5$ and $11$ RBs.


\begin{table}
	\caption{Performance by the combined defense of ``Q-Protect", ``RandomTop", and ``MisNACK" schemes under the RL based attack when there are $5$ RBs.}
	\centering
		{\vspace{0.2cm}
	\footnotesize	\begin{tabular}{c|c|c|c}
Maximum & Recovery & Maximum & Total \\
jammed RBs & time & reduction in reward & reduction \\
\hline \hline
1 & 986 & 0.536 & 266.581 \\ \hline
2 & 918 & 0.679 & 320.966
	\end{tabular}
	}
	\label{table:randomtop5}
\end{table}



\section{Conclusion} \label{sec:conc}
In this paper, we studied the security vulnerability of NextG network slicing by designing a jamming attack on the underlying RL operations for resource allocation.
Although RL is an efficient solution to optimally allocate network resources (RBs at the NextG gNodeB) for communication requests from NextG UEs, the broadcast nature of wireless communications makes the NextG RAN vulnerable to jamming attacks. In particular, if an RB is assigned to a request and is jammed by an adversary, that request cannot be satisfied and the associated reward becomes zero. This reward is used as input to the gNodeB's RL algorithm and thus its performance starts deteriorating. Even after the adversary stops jamming, the gNodeB's performance cannot be recovered until its algorithm is updated by a sufficient number of feedback messages.

To select the RBs for jamming, the adversary builds a surrogate RL model to maximize the number of jammed requests over time subject to an energy budget (namely, a constraint on the number of channels that can be jammed simultaneously). We showed that such an algorithm is highly effective to reduce the gNodeB's performance, even after the adversary stops attacking. We compared this attack with other attack benchmarks such as random jamming and myopic jamming (that aims to maximize the instantaneous number of jammed RBs) and showed that the RL based jamming attack is more effective than both random or myopic jamming.

To protect network slicing against RL based jamming attacks, we introduced several defense schemes such as
suspending
the Q-table updates when an attack is detected, introducing randomness into network slicing decisions or manipulating the feedback mechanism in network slicing to mislead the learning process of the adversary. We showed that these defense schemes can be effectively combined to defend network slicing by fooling the adversary into making wrong decisions and reducing its impact.

\end{document}